\documentclass[final,5p,times,twocolumn]{elsarticle}

\usepackage{amssymb}
\usepackage{bm}
\usepackage{amsmath,graphicx}

\begin{document}
\begin{frontmatter}

\title{Electronic spectral shift of oxygen-filled (6,6) carbon nanotubes}

\author[sh]{Hiroyuki Shima\corref{cor1}}
\author[yo]{Hideo Yoshioka}
\cortext[cor1]{Corresponding Author: shima@eng.hokudai.ac.jp}

\address[sh]{Division of Applied Physics, Faculty of Engineering,
Hokkaido University, Sapporo, Hokkaido 060-8628, Japan}
\address[yo]{Department of Physics, Nara Women's University, Nara 630-8506, Japan}

\begin{abstract}
Electronic state modulation of the armchair (6,6) carbon nanotubes filled with a linear assembly
of oxygen molecules is addressed theoretically.
Ferromagnetic coupling of encapsulated oxygen produces a magnetic field with cylindrical symmetry,
which deviates the electron's eigenenergies from those prior to the oxygen absorption.
An intriguing spectral gap arises near the Fermi energy, at which the gap formation
is allowed only when the tube length equals to a multiple of three in units of carbon hexagon.
A possible means to detect the selective gap formation is discussed.
\end{abstract}


\end{frontmatter}

\section{Introduction}

The central cavity of carbon nanotubes 
is an ideal container for atoms and small molecules.
Initiated by the massive H$_2$-storage observation \cite{DillonNature1997},
many attempts to introduce foreign entities into the hollow space of nanotubes
have been reported: H$_2$O \cite{GordilloCPL2000,ManiwaCPL2005,ManAlexiadisChemRev2008}, 
H$_2${\rm O}$_2$ \cite{RamachCPL2009}, C{\rm O}$_2$ \cite{AlexCPL2008}, C$_2$H$_2$ \cite{GKimCPL2005}, 
and Ar\cite{GaddScience1997} are only a few to mention.
In addition to the potential utility for energy-storage materials and molecular transport devices,
the filling of carbon nanotubes is believed to open up novel applications of drug delivery 
to the cell \cite{RamachCPL2009,HilderNTN2007,ZLiuNanoRes2009}.
In light to mechanical flexibility of nanotubes \cite{ShimaBook}, furthermore,
effects of their cross-sectional deformation on the properties of the encapsulated molecules 
have also been suggested \cite{ShimaNTN2008,ShimaPRB2010}.

Among many kinds of intercalates, special attention has to be paid for 
dioxygen molecules ({\rm O}$_2$) confined into single-walled carbon nanotubes (SWNTs).
An {\rm O}$_2$ molecule is the best known realization of a $p$-electron magnetic material;
it has two unpaired electrons in a doubly degenerate antibonding orbital,
thus being a spin $S=1$ magnet in the ground state.
In the bulk solid state, {\rm O}$_2$ assembly often exhibits anti-ferromagnetic ordering
with mutually parallel arrangement \cite{FreimanPhysRep2004,KlotzPRL2010}.
But confinement into thin long cavity can lead ferromagnetic couplings of adjacent {\rm O}$_2$ molecules,
in which the O-O bond orientations of two entities are orthogonal 
to each other \cite{HemertPRL1983,BusseryJCP1993}.
In fact, the ferromagnetic ordering of a one-dimensional {\rm O}$_2$ chain
is available through confinement into an armchair SWNT of the (6,6) chilarity,
as evidenced by molecular dynamics simulations \cite{HanamiJPSJ2010}.
Structural control of such magnetic molecular aggregates
is attractive in view of low-dimensional physicochemistry,
and thus attempts based on various host materials 
have been made to stabilize the ferromagnetic order of absorbed {\rm O}$_2$ molecules
\cite{TakamizawaJACS2008,RiyadiChemMater2011}.

A possible consequence of the ferromagnetic {\rm O}$_2$ molecular chain inside a SWNT is that
the resulting magnetic field affects electronic properties of 
the host SWNT.
Magnetic field application in general breaks time-reversal symmetry of the system.
This symmetry breaking can lead feasible field-induced phenomena such as
the Aharonov-Bohm effect \cite{AB1,AB2} and the Landau level splitting \cite{Landau}
depending on the field strength and direction.
Particularly in the {\rm O}$_2$-filled (6,6) SWNT,
encapsulated {\rm O}$_2$ molecules are extremely close to the surrounding carbon layer;
therefore, their magnetic moments are thought to yield significant alteration in the quantum states
of $\pi$ electrons,
which has remained unsettled so far.

In this work, we theoretically investigate the electronic structures
of finite-length armchair (6,6) SWNTs encapsulating the ferromagnetic {\rm O}$_2$ chain.
Spectral shifts caused by the {\rm O}$_2$ absorption are found to be strongly dependent
both on the tube length and the energy region to be studied.
A signature of the {\rm O}$_2$ chain confinement manifests in the field-induced gap 
at the Fermi energy.
Interestingly, the gap arises only when the tube length equals to a multiple of three
in units of carbon hexagon, while it vanishes other tube lengths.
This length-dependent gap formation may be observed in the electronic response
of the {\rm O}$_2$-filled nanotubes to alternate voltage with tens gigaheltz frequency.

\section{Model and Methods}

\subsection{Hamitonian}

\begin{figure}[ttt]
\begin{center}
\includegraphics[width=8.5cm]{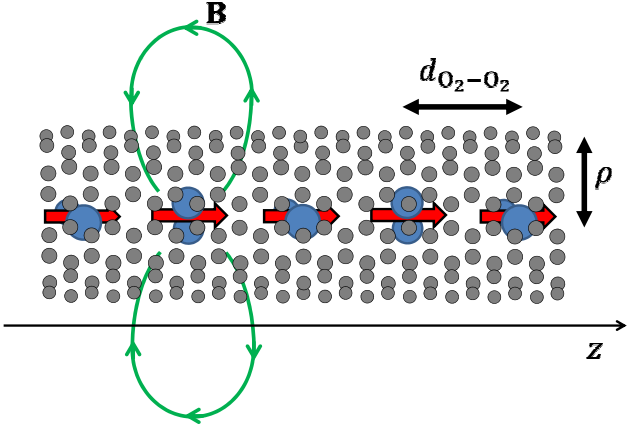}
\end{center}
\caption{Schematic of the {\rm O}$_2$-filled armchair (6,6) SWNT with the tube radius $\rho = 0.407$ nm.
Small (colored in gray) and large (blue) circles are 
the constituent atoms of carbon and oxygen, respectively.
Thick arrows indicate the magnetic dipole moments $\bm{\mu}_{{\rm O}_2}$
associated with the unpaired electrons in {\rm O}$_2$ molecules,
which align with a separation $d_{{\rm O}_2-{\rm O}_2} = 0.33$ nm.
The magnetic field profile $\bm{B}$ generated by single {\rm O}$_2$ molecule 
is depicted by curved paths.}
\label{fig_schematic}
\end{figure}

We consider the electronic structure of {\rm O}$_2$-filled (6,6) SWNTs
utilizing the nearest-neighbor tight-binding model. 
The host graphitic sheet is the one rolled up in the direction of a C-C bond
with the bond length of $a_{\rm C-C} = 0.142$ nm,
having a structural periodicity of the lattice constant of
$a = \sqrt{3} a_{\rm C-C} = 0.246$ nm along the tube axis.
The tube length $L$ is thus written by $N_y a$,
where $N_y$ is the number of hexagons aligned in the axial direction.
Accordingly, $12 N_y$ hexagons composed of $24(N_y+1) [\equiv N_{\rm C}]$ carbon atoms
are involved in the system.
The dangling bonds at the tube edges are assumed to be passivated by hydrogen atoms.

The Hamiltonian matrix is built from the subspace 
spanned by the $N_{\rm C}$ wave functions of 2p$_z$ orbitals. 
The nearest-neighbor Hamiltonian is given by
$\hat{\cal H} = \sum_{i,j} \gamma_{i,j} \hat{c}_i^{\dag} \hat{c}_j$,
where $\gamma_{i,j}$ is the transfer energy between the nearest-neighbor sites, 
and $\hat{c}_i^{\dag} (\hat{c}_i)$ is the creation (annihilation) operator of 
the $\pi$ electron at the site $i$. 
The electronic states are obtained by diagonalizing the $N_C \times N_C$
Hermitian matrix.
When the system is subjected to a magnetic field, 
the transfer integral $\gamma_{i,j}$ becomes $\gamma_{i,j} = \gamma_0 e^{2\pi i \phi_{i,j}}$
with $\gamma_0 = 2.7$ eV, 
where $\phi_{i,j} = (e/h)\int_{\bm{r}_i}^{\bm{r}_j} \bm{A}\cdot d\bm{l}$ is the Peierls phase
and $\bm{A}$ is the vector potential corresponding to the magnetic field.
In the present system, the ferromagnetic {\rm O}$_2$ chain produces a spatially modulated
magnetic field, and the magnetic flux passing through each carbon hexagonal ring
provides a finite value of $\phi_{i,j}$ at edges of the hexagon.

\subsection{Magnetic field distribution}

The stable magnetic spin ($S=1$) in an isolated {\rm O}$_2$ molecule 
generates the magnetic field $\bm{B}_{{\rm O}_2}$ at a point 
$\bm{r} = \rho \bm{e}_{\rho} + z \bm{e}_z$
in terms of the cylindrical polar coordinates $(\rho,\theta,z)$.
The induced magnetic field is written by
\begin{equation}
\bm{B}_{{\rm O}_2}(\bm{r}) 
= \frac{\mu_0}{4\pi} \frac{3\bm{r}(\bm{r}\cdot \bm{\mu}_{O_2}) - r^2 \bm{\mu}_{{\rm O}_2}}{r^5},
\end{equation}
where
$\bm{\mu}_{{\rm O}_2} = -g_e \mu_B \bm{S}/\hbar$ with $g_e = 2$, $\bm{S} = \hbar \bm{e}_z$,
and $\mu_B = e\hbar/(2m_e)$;
$e$ and $m_e$ are the bare electron's charge and mass, respectively,
and $\mu_0$ is the permeability of vacuum.
Our task is to evaluate the normal component of $\bm{B}_{{\rm O}_2}$,
given by
$B_{\perp} = \bm{B}_{{\rm O}_2} \cdot \bm{e}_{\rho}$,
which exerts just on the graphetic sheet.
Using the relations of $\bm{\mu}_{{\rm O}_2} \cdot \bm{e}_{\rho} = 0$
and $\bm{r}\cdot \bm{e}_{\rho}$ with $\rho$ being the tube radius,
the sum of the normal components, $B_{\perp}^{\rm all}(z)$,
associated with all {\rm O}$_2$ molecules encapsulated in the (6,6) tube
is given by
\begin{equation}
B_{\perp}^{\rm all}(z) 
= \frac{-3 \mu_0 \mu_B}{2\pi} \sum_{n=1}^{N_{{\rm O}_2}} 
\frac{\rho (z - \rho - n d_{{\rm O}_2-{\rm O}_2})}
{\left[ \rho^2 + (z-\rho - n d_{{\rm O}_2-{\rm O}_2})^2 \right]^{5/2}}.
\end{equation}
Here, $d_{{\rm O}_2-{\rm O}_2}$ = 0.33 nm
is the separation between adjacent {\rm O}$_2$ molecules \cite{HanamiJPSJ2010},
and $N_{{\rm O}_2}$ is the total number of the {\rm O}$_2$ molecules
that position at $z \in [\rho, L-\rho]$ with equiseparation.
We set $\rho = 0.407$ nm according to the formula
$\rho = a\sqrt{n^2+m^2+nm}/(2\pi)$ for nanotubes with $(n,m)$ chirality.
As a result, the magnetic flux $\Phi(z)$ that penetrates
a hexagonal plaquette at $z$ reads as
$\Phi(z)=\sqrt{3} B_{\perp}^{\rm all} a^2/2$,
and the Peierls phase $\phi_{i,j}$ can be determined via
the relation $\sum_{\rm hexagon} \phi_{i,j} = \Phi(z)/\Phi_0$
with $\Phi_0=h/e$ being flux quantum.

\begin{figure}[ttt]
\begin{center}
\includegraphics[width=8.0cm]{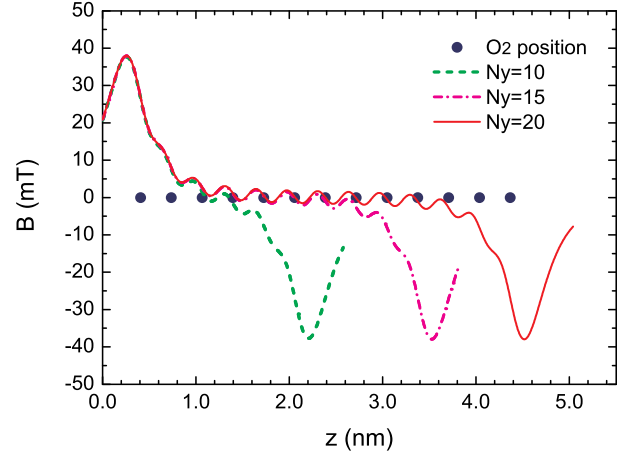}
\end{center}
\caption{Spatial distribution of the normal component $B_{\perp}^{\rm all}$
of the magnetic field caused by the {\rm O}$_2$ molecular chain in the hollow cavity
of (6,6) SWNTs.}
\label{fig_magnetic}
\end{figure}

Figure \ref{fig_magnetic} illustrates the spatial distribution of $B_{\perp}^{\rm all}(z)$
for several tube lengths expressed by $N_y$.
Solid circles indicate the position of {\rm O}$_2$ molecules that may be contained in the host nanotube.
In the $N_y = 10$ case, for instance,
the six {\rm O}$_2$ to the most left ({\it i.e.,} up to the {\rm O}$_2$ at $z=2.06$ nm)
can be included within the allowed region of $z\in [\rho, L-\rho]$ so that $N_{{\rm O}_2} = 6$.
Similarly, $N_{{\rm O}_2} = 10, 13$ for $N_y=15, 20$, respectively.
We observe that $B_{\perp}^{\rm all}(z)$ oscillates with a slight amplitude in-between region
but exhibits a large amplitude on the order of 20 mT near the tube edges.
The amplitude suppression at the intermediate region
is attributed to cancelling out the oppositely-oriented field components
associated with {\rm O}$_2$ molecules separated by $\sim$ 1 nm each other.
Neighborhoods of the tube edges are free from the suppression due to the absence of {\rm O}$_2$ 
both at $z<\rho$ and $z>L-\rho$, and thus significant amount of the Peierls phase is
imposed locally to the electron's eigenmodes.
As the scenario holds regardless of the tube length,
the electronic states for longer (6,6) SWNTs
are modulated to a degree by the {\rm O}$_2$ absorption,
despite of slight magnetic moments
of individual {\rm O}$_2$ molecules.

\section{Results}

\begin{figure}[ttt]
\begin{center}
\includegraphics[width=8.8cm]{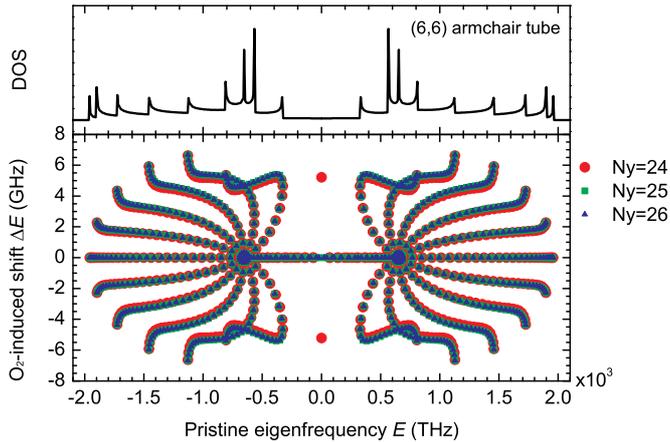}
\end{center}
\caption{
Top: Spectral density of states of an empty ({\it i.e.,} {\rm O}$_2$-unfilled) (6,6) nanotube.
Bottom: {\rm O}$_2$-induced spectral shift for several tube lengths measured by $N_y$.}
\label{fig_dos}
\end{figure}

Figure \ref{fig_dos} displays the {\rm O}$_2$-induced shift in the electronic energy spectra
of short (6,6) SWNTs.
The tube length is set to be $N_y = 24, 25, 26$ as simple examples,
while much longer nanotubes yield similar results as those in this plot.
The horizontal axis $E$ represents the eigenenergies
of the pristine system ({\it i.e.,} prior to {\rm O}$_2$ absorption),
and the vertical axis $\Delta E$ gives their shifts caused by {\rm O}$_2$ absorption.
Each discrete point in the plot indicates the quantum states of isolated finite-length (6,6) SWNTs,
which are written by standing waves of the form $\psi_n(z) \propto \sin(k_n z)$ 
with a discrete set of wavenumbers $k_n=n\pi/(N_y a)$ where $n=1,2,\cdots,N_y$ \cite{VenemaScience1999}.
If the {\rm O}$_2$ chain were absent, $\Delta E \equiv 0$ so that
$N_{\rm C}$ discrete points would be line up horizontally
in the bottom panel of Fig.~\ref{fig_dos}.
But in the current system, the {\rm O}$_2$-induced magnetic field modulates
the $\pi$ electron states,
which results in upward or downward shift in the eigenenergies
as quantified by $\Delta E$ in the figure.

\begin{figure}[ttt]
\begin{center}
\includegraphics[width=7.8cm]{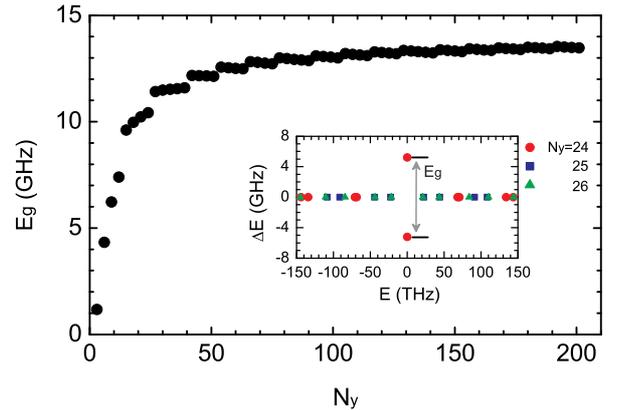}
\end{center}
\caption{HOMO-LUMO energy gap $E_g$ as a function of the tube length $N_y$.
$E_g$ grows monotonically with increasing $N_y$ and converges to $\sim 13.6$ GHz
in the large $N_y$ limit.
Inset: Enlarged view of the {\rm O}$_2$-induced eigenenergy shift in the vicinity of $E=0$.}
\label{fig_gap}
\end{figure}

The data in Fig.~\ref{fig_dos} show pronounced shifts from the pristine values
at discrete energies that correspond to the van Hove singularities
of the spectral density of states (top panel).
A marked exception is the anomalous point-like shift at the band center $(E=E_F=0)$,
which arises only in the case of $N_y = 24$.
This point-like energy shift implies an energy-gap formation between the highest occupied state
and the lowest unoccupied state in the vicinity of $E=0$.
It should be emphasized that
the energy gap no longer disappears in the other two cases of $N_y = 25, 26$
as clarified visually in Inset of Fig.~\ref{fig_gap}.

We have revealed that the {\rm O}$_2$-induced energy gap mentioned above
is universally observed every when ${\rm mod}(N_y, 3) = 0$.
Figure \ref{fig_gap} presents the $N_y$ dependence of the gap magnitude $E_g$;
Note that only $E_g$ at $N_y = 3,6,9,\cdots$ are plotted, 
because $E_g = 0$ whenever ${\rm mod}(N_y,3) \ne 0$.
The result exhibits a monotonic increase with $N_y$ followed by a convergence in the large $N_y$ limit
with the asymptotic value of $E_g \sim 13.6$ GHz.
We mention that $N_y = 201$ corresponds to the tube length $\sim$ 60 nm,
which scale is in a realm of the current fabrication techniques.
It also follows from Fig.~\ref{fig_gap} that the Zeeman splitting is irrelevant
to the gap formation, since the energy scale $\mu_{\rm B} B$ in the present condition
is on the order of 0.1 GHz, far less than the gap magnitude.

\section{Discussion}

\begin{figure}[ttt]
\includegraphics[width=9.0cm]{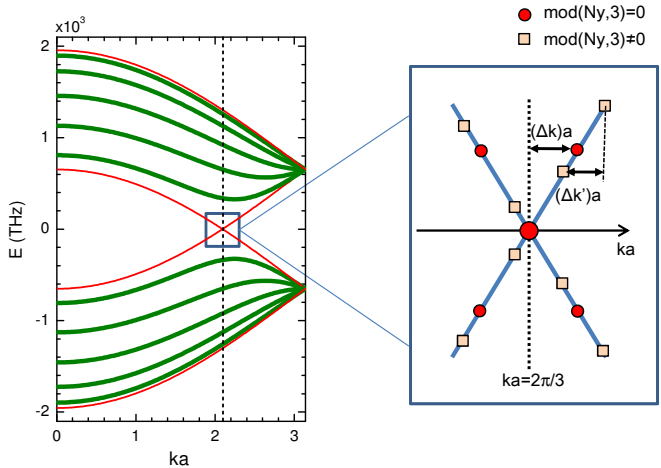}
\caption{
Left: Energy dispersion relation for armchair (6,6) nanotubes
free from the ferromagnetic {\rm O}$_2$ chain.
Ten thick curves are doubly-degenerate at every $k$ points,
while four thin curves are non-degenerate except for the crossing point
at $k=k_F = 2\pi/(3a)$.
Right: Eigenlevel configuration of an empty nanotube near the Fermi energy $E_F = 0$.
Only when ${\rm mod}(N_y, 3) = 0$,
a doubly-degenerate state (indicated by a large solid circle)
resides just on the crossing point.
}
\label{fig_dispersion}
\end{figure}

The length-dependent gap formation at the Fermi level is understood
by considering the eigenlevel configuration for different $N_y$.
The left panel in Fig.~\ref{fig_dispersion} shows 
the energy dispersion relation for the infinitely-long (6,6) nanotube
free from the ferromagnetic {\rm O}$_2$ chain.
There are fourteen branches reflecting the quantum confinement effect
on the $\pi$ electrons along the circumferential direction.
Of these branches, ten thick curves are doubly-degenerate at every $k$ points,
while four thin curves are non-degenerate except for the crossing point
at $k=k_F = 2\pi/(3a)$.
All the degeneracy is lifted when the ferromagnetic {\rm O}$_2$ chain is built
inside the SWNT;
it then leads the spectral shifts of all initially-degenerate eigenmodes that lie 
along the ten thick branches,
which have been already demonstrated in the bottom panel
of Fig.~\ref{fig_dos}.

The abovementioned scenario holds for finite-length (6,6) nanotubes,
while the discreteness of allowed wavenumbers in the axial direction
requires subtle modification.
The right panel of Fig.~\ref{fig_dispersion} gives a drawing
of the level configuration for $N_y$-long {\rm O}$_2$-free (6,6) nanotubes
in the vicinity of $E=0$.
Solid circles indicate the eigenlevels for $N_y$ such that ${\rm mod}(N_y,3) = 0$,
and square ones for ${\rm mod}(N_y,3) \ne 0$.
In the former case, only one doubly-degenerate state (indicated by a large solid circle)
resides just on the crossing point.
In contrast, the latter case involves no degenerate state.
Upon the {\rm O}$_2$ chain encapsulation, therefore,
the degeneracy removal arises only in the $N_y$-long nanotubes with ${\rm mod}(N_y,3) = 0$.
This mechanism accounts for the length-sensitive gap formation at the Fermi level.

The degeneracy at the crossing point owes the availability of an appropriate
value of the discrete wavenumber $k_n =n \Delta k $ with interval $\Delta k = \pi/(N_y a)$.
When $N_y = 3 \ell$ with integer $\ell$,
the interval reads $\Delta k = \pi/(3\ell a)$
and thus the discrete set of $k_n$ can take the value of the Fermi wavenumber $k_F = 2\pi/(3a)$
at $n=2\ell$.
But when $N_y \ne 3\ell$, the resulting interval $\Delta k' = \pi/[(3\ell \pm 1)a]$
forbids any $k_n$ to coincides with $k_F$ since there is no integer $n$
such that $n/(3\ell \pm 1) = 2/3$.

We suggest that the {\rm O}$_2$-induced gap with the magnitude of 13.6 GHz
can be detected through the dynamical conductivity measurement at low temperature.
According to the Kubo formula \cite{Kubo1,Kubo2}, the dynamical conductance 
along the tube axis is found as
\begin{equation}
G(\omega) = -\frac{1}{i\omega}
\sum_{\mu,\nu} | J_{\mu \nu}^z |^2
\frac{f_{\mu}-f_{\nu}}{\hbar\omega + E_{\mu} - E_{\nu}+i\delta},
\label{eq_003}
\end{equation}
where $f_{\mu}$ and $f_{\nu}$ are Fermi distribution functions and 
$\delta$ is positive infinitesimal.
The magnitude of $G(\omega)$ is primarily determined by 
$J_{\mu \nu}^z = \langle E_{\mu} | \hat{J}_z | E_{\nu} \rangle$,
in which $\hat{J}_z$ is the $z$ component of the current operator
$\hat{\bm{J}} = (-ie/\hbar) [\hat{\bm{r}}, \hat{\cal H}]$ with a commutator $[\hat{\cdots}, \hat{\cdots}]$.
Suppose that ac electric voltage with frequency $\omega$
is applied to the {\rm O}$_2$-filled nanotubes.
A sufficiently low temperature of $k_{\rm B} T \le \hbar \omega$,
Eq.~(\ref{eq_003}) is reduced to
$G(\omega) \propto |J_{+-}|^2 /\omega$,
where $J_{+-} = \langle E_+ | \hat{J} | E_- \rangle$
and $E_{+(-)}$ is the eigenlevel immediately above (below) $E_F=0$.
In the present system, $J_{+-}$ has a modestly large value,
which follows from the numerical result of 
$\left| 
\sum_{i=1}^{N_{\rm C}} \langle E_+ |\bm{r}_i \rangle \langle \bm{r}_i + \bm{e}_z |E_- \rangle 
\right| \sim 0.2$
under the normalization condition of 
$\sum_{i=1}^{N_{\rm C}} \left| \langle \bm{r}_i | E_{+(-)} \rangle \right|^2 = 1$.
Hence, the dynamical conductivity will exhibit a sharp peak at $\omega \sim$ 13.6 GHz
in a low-temperature measurement at less than 1 K.
A large-scale computation technique for the Kubo-formula \cite{ShimaPRB1999,ShimaPRB2004}
incorpolated with {\it ab initio} calculations \cite{Umeno2004}
enable a quantitative examination of the conjecture,
which we will perform in the future work.

\section{Summary}

We have theoretically considered the electronic spectral shift 
of the (6,6) SWNT encapsulating a linear ferromagnetic {\rm O}$_2$ chain
into the hollow cavity.
{\rm O}$_2$ absorption causes an energy gap between the highest and lowest occupied states 
only in the $N_y$-long (6,6) SWNTs with $N_y$ being multiples of three;
the gap vanishes for other $N_y$.
The length-sensitive gap formation is attributed to the removal of eigenlevel degeneracy
at the Fermi level that arises only when ${\rm mod}(N_y,3) =0$.
The gap magnitude approaches 13.6 GHz in the long-$N_y$ limit,
which suggests the possibility that the {\rm O}$_2$-induced gap can be confirmed 
in the alternative-current measurement of the host SWNT.

\section*{Acknowledgment}
Fruitful discussion with H.~Suzuura is greatly acknowledged.
This work was supported by MEXT and Nara Women's University Intramural Grant for Project Research. 
HS cordially thanks the financial supports by 
the Inamori Foundation and the Suhara Memorial Foundation.

%
%


\begin{thebibliography}{99}

\bibitem{DillonNature1997}
A.~C.~Dillon, K.~M.~Jones, T.~A.~Bekkedahl, C.~H.~Kiang, D.~S.~Bethune, M.~J.~Heben,
Nature 386 (1997) 377.

\bibitem{GordilloCPL2000}
M.~C.~Gordillo, J.~Mart\'i, Chem.~Phys.~Lett. 329 (2000) 341.

\bibitem{ManiwaCPL2005}
Y.~Maniwa, H.~Kataura, M.~Abe, A.~Udaka, S.~Suzuki, Y.~Achiba, H.~Kira, K.~Matsuda, H.~Kadowaki and Y.~Okabe,
Chem.~Phys.~Lett. 401 (2005) 534.

\bibitem{ManAlexiadisChemRev2008}
A.~Alexiadis, S.~Kassinos, Chem.~Rev. 108 (2008) 5014.

\bibitem{RamachCPL2009}
C.~N.~Ramachandran, D.~D.~Fazio, N.~Sathyamurthy, V.~Aquilanti,
Chem.~Phys.~Lett. 473 (2009) 146.

\bibitem{AlexCPL2008}
A.~Alexiadis and S.~Kassinos,
Chem.~Phys.~Lett. 460 (2008) 512.

\bibitem{GKimCPL2005}
G.~Kim, Y.~Kim and J.~Ihm,
Chem.~Phys.~Lett. 415 (2005) 279.

\bibitem{GaddScience1997}
G. E. Gadd, M. Blackford, S. Moricca, N. Webb, P. J. Evans, A. M. Smith, G. Jacobsen, 
S. Leung, A. Day and Q. Hua,
Science 277 (1997) 933.

\bibitem{HilderNTN2007}
T.~A.~Hilder, J.~M.~Hill,
Nanotechnology 18 (2007) 275704.

\bibitem{ZLiuNanoRes2009}
Z.~Liu, S.~Tabakman, K.~Welsher, H.~Dai,
Nano Res. 2 (2009) 85.

\bibitem{ShimaBook}
H.~Shima, M.~Sato, {\it Elastic and Plastic Deformation of Carbon Nanotoubes},
Pan Stanford Publishing, Singapore (2011).

\bibitem{ShimaNTN2008}
H.~Shima, M.~Sato, Nanotechnology 19 (2008) 495705.

\bibitem{ShimaPRB2010}
H.~Shima, M.~Sato, K.~Iibishi, S.~Ghosh, M.~Arroyo, Phys. Rev. B 82 (2010) 085401.

\bibitem{FreimanPhysRep2004}
Y.~A.~Freiman, H.~J.~Jodl,  Phys.~Rep.~ 401 (2004) 1.

\bibitem{KlotzPRL2010}
S. Klotz, T.~Str\"assle, A.~L.~Cornelius, J.~Philippe, T.~Hansen, 
Phys.~Rev.~Lett. 104 (2010) 115501.

\bibitem{HemertPRL1983}
M.~C.~van~Hemert, P.~E.~S.~Wormer, A.~van~der~Avoird, Phys.~Rev.~Lett. 51 (1983) 1167.

\bibitem{BusseryJCP1993}
B.~Bussery, P.~E.~S.~Wormer, J. Chem. Phys. 99 (1993) 1230.

\bibitem{HanamiJPSJ2010}
K.~Hanami, T.~Umesaki, K.~Matsuda, Y.~Miyata, H.~Kataura, Y.~Okabe, Y.~Maniwa,
J. Phys. Soc. Jpn.79 (2010) 023601.

\bibitem{TakamizawaJACS2008}
S.~Takamizawa, E.~Nakata, T.~Akatsuka, C.~Kachi-Terajima, R.~Miyake, 
J.~Am.~Chem.~Soc. 130 (2008) 17882.

\bibitem{RiyadiChemMater2011}
S.~Riyadi, S.~Giriyapura, R.~A.~de~Groot, A.~Caretta, P.~H.~M.~van~Loosdrecht,
T.~T.~M.~Palstra, G.~R.~Blake,
Chem.~Mater. 23 (2011) 1578.

\bibitem{AB1}
H.~Ajiki, T.~Ando, J.~Phys.~Soc.~Jpn. 62 (1993) 2470.

\bibitem{AB2}
A.~Bachtold, C.~Strunk, J.~P.~Salvetat, J.~M.~Bonard, L.~Forr\'o, 
T.~Nussbaumer, C.~Sch\"onenberger, Nature 397 (1999) 673.


\bibitem{Landau}
H.~Ajiki, T.~Ando, J.~Phys.~Soc.~Jpn. 65 (1996) 505.

\bibitem{VenemaScience1999}
L.~C.~Venema, J.~W.~G.~Wild\"oer, S.~J.~Tans, J.~W.~Janssen, L.~J.~Hinne, T.~Tuinstra, 
L.~P.~Kouwenhoven, C.~Dekker, Science 283 (1999) 52.

\bibitem{Kubo1}
R.~Kubo, J.~Phys.~Soc.~Jpn. 12 (1957) 570.

\bibitem{Kubo2}
U.~Brandt, M.~Moraweck, J.~Phys.~C: Solid State Phys. 15 (1982) 5255.

\bibitem{ShimaPRB1999}
H.~Shima, T.~Nakayama, Phys.~Rev.~B 60 (1999) 14066.

\bibitem{ShimaPRB2004}
H.~Shima, T.~Nomura, T.~Nakayama, Phys.~Rev.~B 70 (2004) 075116.

\bibitem{Umeno2004}
Y.~Umeno, T.~Kitamura, A.~Kushinma, Comp.~Materi.~Sci. 30 (2004) 283.


\end{thebibliography}
\end{document}